\documentclass[aps,preprint,nofootinbib,superscriptaddress]{revtex4}
\usepackage{color}
\usepackage{longtable}
\usepackage{graphicx}
\usepackage{amsmath}
\usepackage{amsfonts}
\usepackage{amssymb}
\usepackage{natbib} 
\usepackage{setspace}
\usepackage{xspace}
\usepackage{cancel}
\usepackage{physics}
\usepackage{tensor,braket,color,bbold,dsfont,slashed}
\usepackage{mathrsfs}
\usepackage[normalem]{ulem}
\begin{document}
\title{Neutrino propagation in the neutron star with uncertainties\\
 from nuclear, hadron, and particle physics}
\author{Parada~T.~P.~Hutauruk}\email{phutauruk@gmail.com}
\affiliation{Department of Physics, Pukyong National University (PKNU), Busan 48513, Korea}
\affiliation{Department of Physics Education, Daegu University, Gyeongsan 38453, Korea}
\author{Hana Gil}\email{khn1219@gmail.com}
\affiliation{Center for Extreme Nuclear Matters (CENuM), Korea University, Seoul 02841, Korea}
\author{Seung-il Nam}\email{sinam@pknu.ac.kr}
\affiliation{Department of Physics, Pukyong National University (PKNU), Busan 48513, Korea}
\affiliation{Center for Extreme Nuclear Matters (CENuM), Korea University, Seoul 02841, Korea}
\affiliation{Asia Pacific Center for Theoretical Physics (APCTP), Pohang 37673, Korea}
\author{Chang Ho Hyun}\email{hch@daegu.ac.kr}
\affiliation{Department of Physics Education, Daegu University, Gyeongsan 38453, Korea}
\affiliation{Center for Extreme Nuclear Matters (CENuM), Korea University, Seoul 02841, Korea}
\date{\today}
\begin{abstract}
In the present work, we investigate the neutral-current neutrino-nucleon scattering in the nuclear medium using various energy-density functional (EDF) models such as the KIDS (Korea-IBS-Daegu-SKKU) and SLy4, together with the quark-meson coupling (QMC) model for the nucleon form factors at finite density. The differential cross section (DCS) and neutrino mean free path (NMFP) are computed numerically, considering the density-dependent nucleon form factors (DDFF) and neutrino structural properties such as the neutrino magnetic moment (NMM) and its electric charge radius (NCR). It turns out that the DDFF decreases the scattering cross-section, while the NCR increases it considerably. The effect of the NMM turns out to be almost negligible. We also observe that the value of the neutron effective mass is of importance in the neutron-star cooling process, indicating that for the neutron effective mass larger than the mass in free space, the neutrino can interact with matter at densities $\rho \gtrsim 1.5 \rho_0$ in the neutron star with radius 13 km.
\end{abstract}
\maketitle
\section{Introduction} \label{sec:intro}
It is widely known that the absorption and scattering of a neutrino with matter constituents play crucial roles in the evolution of stellar collapse and the success or failure of the supernova explosion. It is also known that the cooling of young neutron stars is driven by the emission of neutrinos via processes such as direct URCA, modified URCA, and neutrino bremsstrahlung. Neutrinos that are created in these processes can either escape from the neutron star freely or scatter inelastically or even be trapped before they reach the surface of the neutron star. If neutrinos are trapped by the neutron star matter, it can delay the rate of the cooling process, and as a consequence give a significant effect on the cooling curve of the neutron star.

In the recent work, we investigated the weak interaction of neutrinos in the homogeneous neutron-star (NS) matter within the framework of Korea-IBS-Daegu-SKKU (KIDS) density functional~\cite{Hutauruk:2022bii}. The work focused on the effect of uncertainties and/or corrections of the nuclear matter equation of state (EoS), i.e., symmetry energy and nucleon effective mass, to the neutrino mean free path (NMFP) within the NS systematically. We found that the NMFP depends strongly on these uncertainties and/or corrections. Compared with the NS radius, the NMFP could be as short as about half of the NS radius but also could be larger than the NS radius. Such wide-range values of NMFP can lead to a different result in the cooling behavior of the NS. Overall, the work of Ref.~\cite{Hutauruk:2022bii} demonstrated the importance of the accurate determination of nuclear matter EoS in the neutrino-weak interaction at finite density and zero temperature. However, it is worth noting that in Ref.~\cite{Hutauruk:2022bii}, the nucleons are treated as point particles, while, in fact, the nucleon has a structure such as the electromagnetic form factor in the transverse momentum and parton distribution functions in the longitudinal momentum. Such structure of the nucleon has been confirmed not only theoretically~\cite{Miller:2007kt} but also experimentally~\cite{Hyde:2004gef,Andivahis:1994rq,Litt:1969my,SAMPLE:1997dds}. Moreover, in the past, many studies have been done in investigating the neutral and charged current weak interaction of a neutrino with the matter by considering the electromagnetic form factor in the free space~\cite{Horowitz:2003yx,Reddy:1998hb,Sulaksono:2005wv,Guo:2020tgx,Sulaksono:2006eu,Hutauruk:2006re}. In the present work, we consider the structure of the nucleon--which is originally from non-perturbative quantum chromodynamic (QCD) aspect of hadron and particle physics in both free space and medium. It is well known that the electromagnetic form factors are expected to be modified in the medium~\cite{Cloet:2009tx,Lu:1998tn,Geesaman:1995yd,EuropeanMuon:1983wih,Malace:2008gf,Hutauruk:2018cgu,Hutauruk:2020mhl}. Such medium effect can be non-negligible in the model~\cite{plb2009,jpg2010}, providing another source of the corrections that can affect the NMFP.

In most standard studies, neutrinos are assumed to be elementary particles. However, several experiments show the non-zero values for the magnetic moment and charge radius of the neutrino~\cite{Super-Kamiokande:2004wqk,TEXONO:2002pra,MUNU:2003peb,Beda:2013mta,XENON:2020rca,Borexino:2017fbd,Allen:1992qe,Cadeddu:2018dux}, although the evidence is not firmly established yet. More theoretical studies and experiments using modern facilities like CONUS~\cite{CONUS:2022qbb}, DUNE~\cite{Jana:2022tsa}, NOMAD~\cite{NOMAD:2009qmu}, MiniBooNe~\cite{MiniBooNE:2010xqw}, MINERvA~\cite{MINERvA:2013kdn}, Hyper-Kamiokande, and other reactor or accelerator experiments as well as galactic or atmospheric neutrinos are really needed to collect more data in order to establish the properties of the neutrino. If the neutrinos have an internal structure, this evidence of neutrino moment magnetic (NMM) and charge radius (NCR) will significantly impact elementary particle physics, nuclear physics, astrophysics, and cosmology in the standard model calculation.

In this study, we take into account the aforementioned uncertainties and/or corrections from nuclear physics, hadron physics, elementary particle physics, i.e., nucleon form factor (free space and medium), and NMM and NCR in the description of neutrino electro-weak interaction at finite density and zero temperature. The values of the NMM and NCR are obtained from experimental constraints~\cite{Beda:2013mta,Allen:1992qe,Cadeddu:2018dux}. We then calculate the differential cross-section (DCS) for the neutrino-nucleon scattering and NMFP of the neutrinos in the core of the NS. The roles of the nucleon and neutrino corrections in the neutrino propagation inside NS are explored in further detail. We find that the in-medium corrections of the nucleon form factor and NCR give a significant impact on the DCS and NMFP. Compared to the vacuum nucleon form factor (VFF), the density dependence of the nucleon form factor (DDFF) decreases the DCS, implying an increase in the NMFP. The increase of the NMFP is more significant at higher matter density. It means that the neutrino with the DDFF will more freely escape from the core of NS. The DCS does not change significantly by taking an NMM value from the experimental constraints~\cite{Beda:2013mta}. This indicates that the NMM plays a small role in the neutrino emission from the NS core. However, it is drastically changed when we consider a finite NCR value obtained from the experiment constraint~\cite{Allen:1992qe,Cadeddu:2018dux}. It increases the DCS significantly in comparison with other scenarios without the NCR. This shows that the neutrinos are more strongly interacting with the matter if the NCR contribution is taken into account appropriately. 

The present work is organized as follows: Section II is devoted to a brief description of the theoretical framework. The numerical results and related discussions are given in Section III. The final Section is devoted to the summary.

\section{Neutrino-nucleon interaction} \label{sec:ermf}
Here we adopt four non-relativistic energy density functional (EDF) models: KIDS0, KIDS-A, KIDS0-m*87, and SLy4~\cite{Hutauruk:2022bii}. All the models have yielded identical values of the saturation density $\rho_0=0.16$~fm$^{-3}$ and the binding energy per nucleon $E_{\rm B}=16$~MeV. However, they have distinctive behavior for the EoS at densities below and above the saturation. A conventional expansion of the energy per nucleon can be written as
\begin{eqnarray}
  \label{eq1}
        {\cal E}(\rho,\, \delta) &=& E(\rho) + S(\rho) \delta^2 + O(\delta^3), \\
        E(\rho) &=& E_{\rm B} + \frac{1}{2} K_0 x^2 + O(x^3),  \\
        S(\rho) &=& J + L x + \frac{1}{2}K_{\rm sym} x^2 + O(x^3),\,\,\,\,\,\, x = \frac{\rho - \rho_0}{3 \rho_0},\,\, \delta = \frac{\rho_n - \rho_p}{\rho}.
\end{eqnarray}
%
\begin{table}
  \begin{center}
    \begin{tabular}{|c|c|c|c|c|c|c|}\hline
      & $K_0$ & $J$ & $L$ & $K_{\rm sym}$ & $m^*_s/M$ & $m^*_v/M$ \\ \hline
      KIDS0 & 240 & 32.8 & 49.1 & $-156.7$ & 1.0 & 0.8 \\ 
      KIDS-A & 230 & 33 & 66 & $-139.5$ & 1.0 & 0.8 \\
      KIDS0-m*87 & 240 & 32.8 & 49.1 & $-156.7$ & 0.8 & 0.7 \\
      SLy4 & 229.9 & 32 & 45.9 & $-119.7$ & 0.7 & 0.8 \\
      \hline
    \end{tabular}
  \end{center}
  \caption{Nuclear matter parameters of the models.
    $K_0$, $J$, $L$ and $K_{\rm sym}$ are in the units of MeV.
    $m^*_s$, $m^*_v$ and $M$ denote the isoscalar effective mass, isovector effective mass, and nucleon mass in free space,
    respectively.}
  \label{tab1}
\end{table}
%
The parameters that characterize the density-dependence of EoS are summarized in Tab.~\ref{tab1}. The KIDS0, KIDS0-m*87, and SLy4 models have similar density dependencies for the symmetry energy but are very different in the effective mass. On the other hand, the KIDS0 and KIDS-A models have similar effective masses, whereas the symmetry energy for the KIDS-A model is much stiffer than that for the KIDS0 model. Comparison among these models will show the role of the effective mass and symmetry energy in the neutrino-nucleon interaction in the nuclear medium.

The neutral-current weak and electromagnetic (EM) interactions for the neutrino-nucleon scattering in the nuclear medium can be described in terms of the following  effective Lagrangian:
\begin{eqnarray}
  \label{eq2}
        {\mathscr L}^N_{\rm int} &=& \frac{G_F}{\sqrt{2}} \left(\bar{\nu} \Gamma^\mu_{\rm W} \nu \right)
        \left( \bar{N}   J^{\rm W}_\mu  N \right) 
        + \frac{4 \pi \alpha_{\rm EM}}{q^2} \left(\bar{\nu}\Gamma^\mu_{\rm EM} \nu\right)
        \left(\bar{N}   J^{\rm EM}_\mu  N\right),
\end{eqnarray}
where the $\nu$ and $N=(n,p)$ denote the neutrino and nucleon fields, respectively. The weak and EM currents for the nucleon, $J^{\rm W}_\mu$ and $J^{\rm EM}_\mu$ are defined by
\begin{eqnarray}
  \label{eq3}
  J^{\rm W}_\mu &=& F^{\rm W}_1(q^2) \gamma_\mu - G_A(q^2) \gamma_\mu \gamma^5
  + i F^{\rm W}_2(q^2)\frac{\sigma_{\mu \nu} q^\nu}{2M} + \frac{G_p(q^2)}{2M} q_\mu \gamma^5, \nonumber \\
  J^{\rm EM}_\mu &=& F^{\rm EM}_1(q^2) \gamma_\mu + i F^{\rm EM}_2(q^2)\frac{\sigma_{\mu\nu}q^\nu}{2M}.
\end{eqnarray}
The values of the nucleon form factors $G_A$ and $F_{1,2}^W$ at $q^2 =0$ in vacuum are summarized in Tab.~\ref{tab2}.
\begin{table}
  \begin{center} 
  \begin{tabular}{|c|c|c|c|c|c|}\hline
    Target & $G_A$ & $F^{\rm W}_1$ & $F^{\rm W}_2$ & $F^{\rm EM}_1$ & $F^{\rm EM}_2$ \\ \hline
    $n$ &$-\frac{g_A}{2}$  & $-0.5$  & $-\frac{1}{2}(\kappa_p -\kappa_n) - 2 \sin^2 \theta_w \kappa_n$ & 0 & $\kappa_n$ \\
    $p$ & $\frac{g_A}{2}$ & $0.5 - 2 \sin^2 \theta_w$ & $\frac{1}{2}(\kappa_p - \kappa_n)-2 \sin^2 \theta_w \kappa_n$ & 1 & $\kappa_p$ \\
    \hline
  \end{tabular}
\end{center}
\caption{Vacuum form factor values at $q^2=0$. In the numerical calculation, we use $\sin^2\theta_w = 0.231$, $g_A=1.260$, $\kappa_p=1.793$ and $\kappa_n=-1.913$.}
\label{tab2}
\end{table}

Since we are interested in the density effects of the neutrino-nucleon scattering, the nucleon form factors should be described as functions of density. For this purpose, the DDFFs are calculated in the quark-meson coupling (QMC) model~\cite{Hutauruk:2018cgu}. The QMC model is built in the quark degree of freedom and has been successfully and widely used in many applications of physics phenomena such as properties of hadron, neutron star, and properties of finite nuclei. In the application to nuclear properties, the QMC model reproduces the nuclear charge distribution of shell nuclei, saturation energy, and compressibility of nuclear matter at high accuracy. For the hadron properties in nuclear medium, the QMC model was used to calculate the DDFFs at the quark level, where so far, not many calculations for DDFFs at the quark level are available in the literature

The weak-interaction vertex of the Dirac neutrino in Eq.~(\ref{eq2}) can be written  in the standard $V-A$ form as follows:
\begin{equation}
  \Gamma^\mu_{\rm W} = \gamma^\mu (1-\gamma^5),
\end{equation}
while the EM-interaction vertex is constructed in terms of the four independent form factors in general
\begin{eqnarray}
\label{GAMMAEM}
  \Gamma^\mu_{\rm EM} = f_1 \gamma^\mu - \frac{i}{2m_e} f_2\sigma^{\mu\nu}q_\nu
  + g_1\left(g^{\mu\nu}-\frac{q^\mu q^\nu}{q^2} \right)\gamma_\nu \gamma^5
  -\frac{i}{2m_e} g_2\sigma^{\mu\nu}q_\nu \gamma^5.
\end{eqnarray}
Here, $f_1$, $g_1$, $f_2$, and $g_2$ are called the Dirac, anapole, magnetic, and electric dipole
form factors as functions of $q^2$, respectively. The NCR is simply defined by
\begin{eqnarray}
  \left< R^2_\nu \right> \equiv \left<R^2_V \right> + \left< R^2_A \right>,
\end{eqnarray}
where $R_V$ and $R_A$ are the vector and axial-vector charge radii, which are defined by
\begin{eqnarray}
  \left< R^2_V \right> &=& \left. 6 \frac{d f_1(q^2)}{d q^2} \right|_{q^2=0},\,\,\,\,\left< R^2_A \right> = \left .6 \frac{d g_1(q^2)}{d q^2} \right|_{q^2=0}.
\end{eqnarray}
By doing this, we can explore the effects of the spatial extension of the neutrino in the medium. In the Breit frame with $q_0=0$, we can use the approximate relation:
\begin{eqnarray}
  f_1(q^2) &\simeq& - \frac{1}{6} \left< R^2_V \right> \mathbf{q}^2,\,\,\,\,g_1(q^2) \simeq - \frac{1}{6} \left< R^2_A \right> \mathbf{q}^2.
\end{eqnarray}
At $q^2=0$, $f_2(q^2)$ and $g_2(q^2)$ define the NMM and charge-parity violating electric dipole moment as
\begin{equation}
\mu^m_\nu = f_2(0) \mu_B \,\,\,\,\, {\rm and} \,\,\,\,\, \mu^e_\nu = g_2(0) \mu_B,
\end{equation}
from which we can define the effective NMM as
\begin{equation}
\mu^2_\nu \equiv (\mu^m_\nu)^2 + (\mu^e_\nu)^2
\end{equation}
with the Bohr magneton $\mu_B = \frac{e}{2m_e}$, where $e$ and $m_e$ is the electron unit charge and mass, respectively.

The DCS density for the neutrino-nucleon scattering in the weak and EM neutral-current interactions is given by
\begin{eqnarray}
\label{eq4}
\frac{1}{V} \frac{d^3\sigma}{dE'_\nu \, d^2\Omega} &=&   - \frac{1}{16\pi^2} \frac{E'_\nu}{E_\nu} 
\Big[ \frac{G_F^2}{2} \left(L_\nu^{\alpha \beta} \Pi_{\alpha \beta}^{\mathrm{Im}} \right)_{\mathrm{W}} 
+ \frac{16 \pi^2 \alpha_{\mathrm{EM}^2}}{q^4} \left( L_\nu^{\alpha \beta} \Pi_{\alpha \beta}^{\mathrm{Im}} \right)_{\mathrm{EM}} 
\nonumber \\    
&+& \frac{8\pi G_F \alpha_{\mathrm{EM}}}{q^2 \sqrt{2}} \left( L_\nu^{\alpha \beta} \Pi_{\alpha \beta}^{\mathrm{Im}} \right)_{\mathrm{INT}} \Big]
\end{eqnarray}
where $E'_\nu$ and $E_\nu$ are respectively the final and initial neutrino energies. The detailed forms of $L_\nu$ and $\Pi^{\rm Im}$ for the weak, EM and interference terms are given in Refs.~\cite{Reddy:1998hb,Sulaksono:2006eu,Hutauruk:2018cgu}.

The inverse of the neutrino mean free path (NMFP) is determined by integrating the DCS in Eq.~(\ref{eq4}) over $q_0$ and $|\textbf{q}|$, resulting in
\begin{eqnarray}
  \label{eq5}
  \lambda^{-1} (E_\nu) &=& 2 \pi \int_{q_0}^{{(E_\nu + E_\nu')}} d|\mathbf{q}| \int_0^{2E_\nu} dq_0 \frac{|\mathbf{q}|}{E_\nu E_\nu'}  \left[ \frac{1}{V} \frac{d^3 \sigma}{dE_\nu' d^2\Omega} \right].
\end{eqnarray}
Here, $E_\nu' = E_\nu -q_0$.

\section{Numerical result and discussion} \label{sec:NRCC}
In this Section, we present the numerical results with detailed discussions of the DDFF, DCR, and NMFP. 

\subsection{Nucleon DDFF}
\begin{figure}[t]
\begin{tabular}{ccc}
\includegraphics[width=5cm]{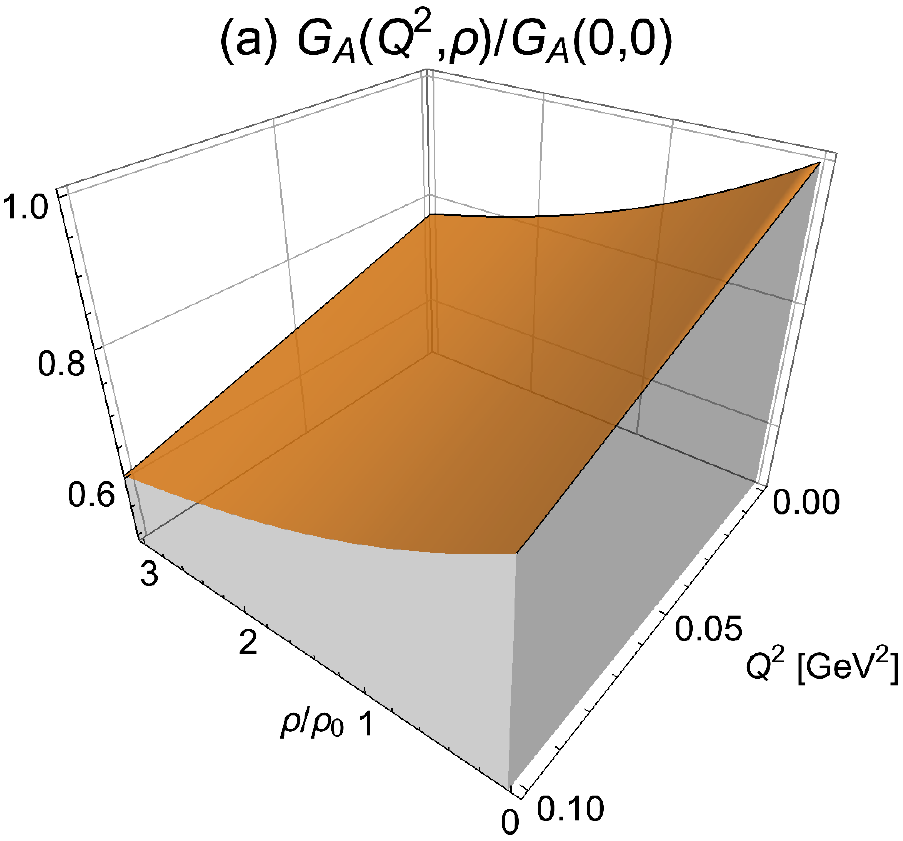}
\includegraphics[width=5cm]{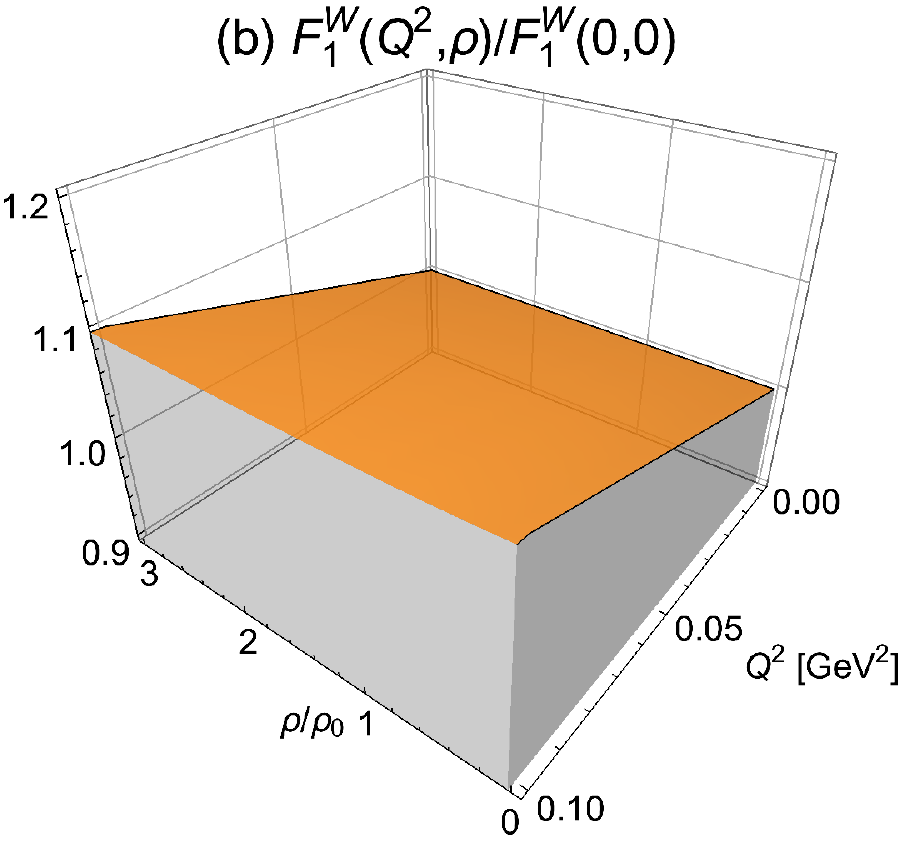}
\includegraphics[width=5cm]{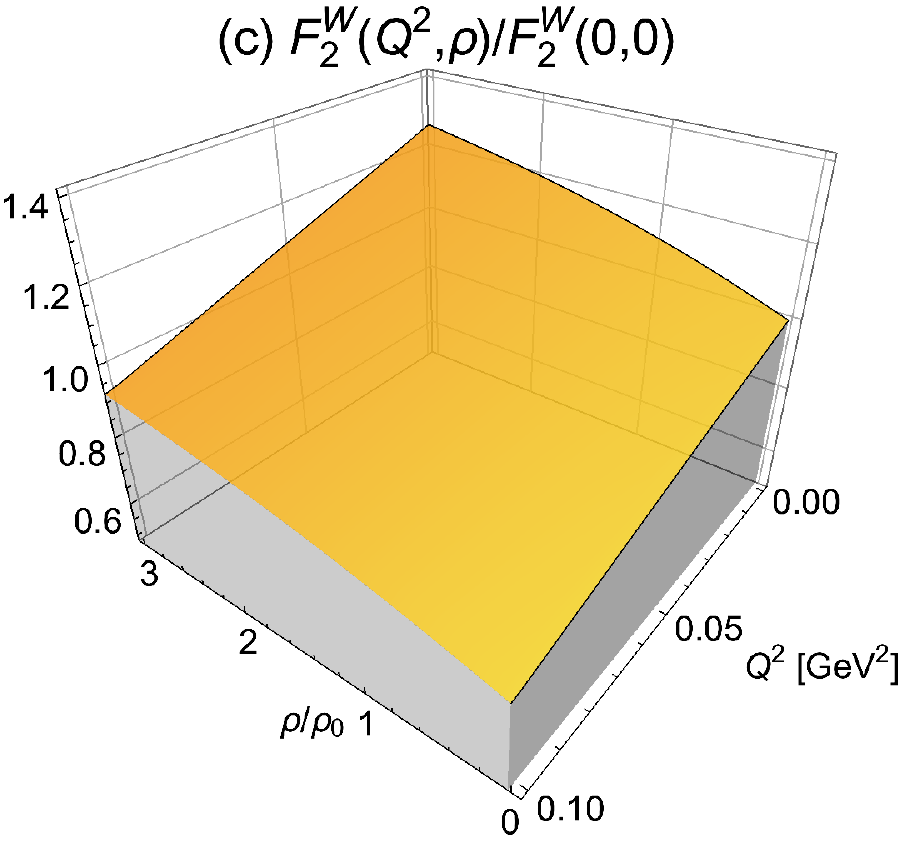}
\end{tabular}
 \caption{Normalized density-dependent weak form factors (DDFF) for the nucleon as functions of $Q^2\le 0.1\,\mathrm{GeV}^2$ and $\rho/\rho_0\le 3$ 
 from the QMC model~\cite{Saito:2005rv}: (a) $G_A(Q^2,\rho)/G_A(0,0)$, (b) $F^W_{1}(Q^2,\rho)/F^W_{1}(0,0)$, and (c) $F^W_{2}(Q^2,\rho)/F^W_{2}(0,0)$.}
  \label{fig1}
  \end{figure}

As mentioned already, since we are interested in the neutrino-nucleon scattering inside the nuclear medium, density effects should be taken into account carefully for the nucleon form factors. For this purpose, we employ the QMC model~\cite{Saito:2005rv}. In Fig.~\ref{fig1}, we depict the numerical results for the normalized density-dependent weak form factors for the nucleon as functions of $Q^2$ and $\rho/\rho_0$, showing (a) $G_A(Q^2,\rho)/G_A(0,0)$, (b) $F^W_{1}(Q^2,\rho)/F^W_{1}(0,0)$, and (c) $F^W_{2}(Q^2,\rho)/F^W_{2}(0,0)$ for the proton. Note that the $F^W_1$ is almost independent of the density in the low $Q^2$ region, in which we are interested mostly, whereas the $G_A$ decreases with respect to the density, as shown in panel (a) of Fig.~\ref{fig1}. This decreasing behavior of the $G_A$ can be understood by the lower component of the quark spinor $\mathcal{L}(r)\propto\mathcal{O}(1/M^*_q)$ is enhanced more than its upper component $\mathscr{U}(r)\propto\mathcal{O}(1)$ as functions of the density when it is calculated using the three-point quark operator~\cite{Saito:2005rv,Lu:2001mf}: $G_A \equiv \int d^3r \left[ \mathscr{U}^2(r) -\mathcal{L}^2(r)\right]>1$. On the contrary, the $F_2^W$ increases with respect to the density, and this tendency is originated from $F_2^W \equiv \int d^3r  \mathscr{U}(r) \mathcal{L}(r)$. From the numerical calculations, we verified that the contributions from the $G_A$ are the most dominant to describe the scattering cross-section. Hence, with the DDFF, the scattering cross-section will be reduced with respect to the density in comparison to that without it. It is worth mentioning that the result for the $G_A$ at saturation density, for instance, turns out to be consistent with other theoretical calculations~\cite{Hutauruk:2018qku,Lu:2001mf,Rakhimov:1998hu}. Although we do not show the EM form factors here, we verify that their density dependencies are not significant in the cross-section.

\subsection{Neutrino-nucleon scattering DCS}
\begin{figure}[t]
\begin{tabular}{cc}
\includegraphics[width=8cm]{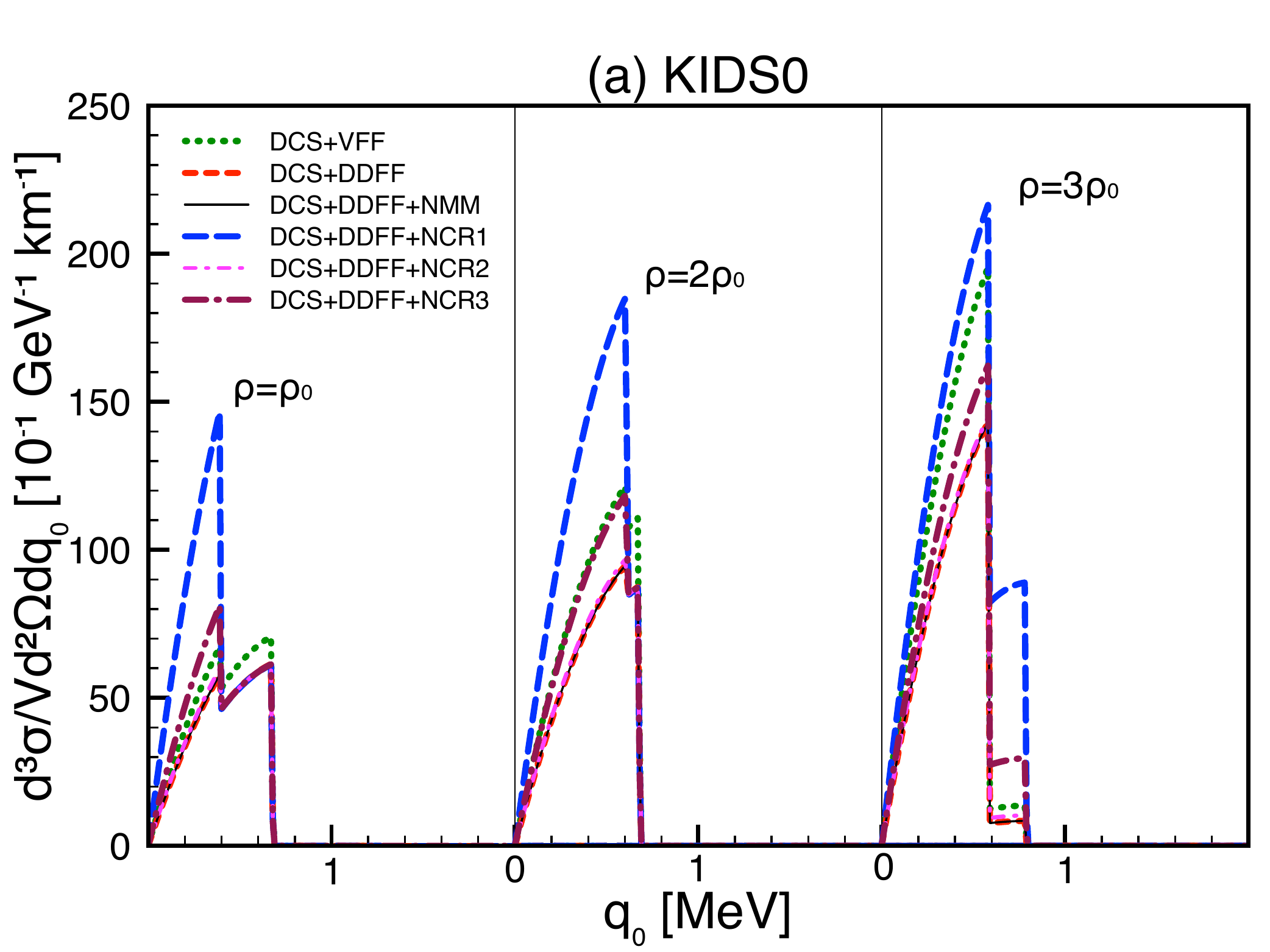}
\includegraphics[width=8cm]{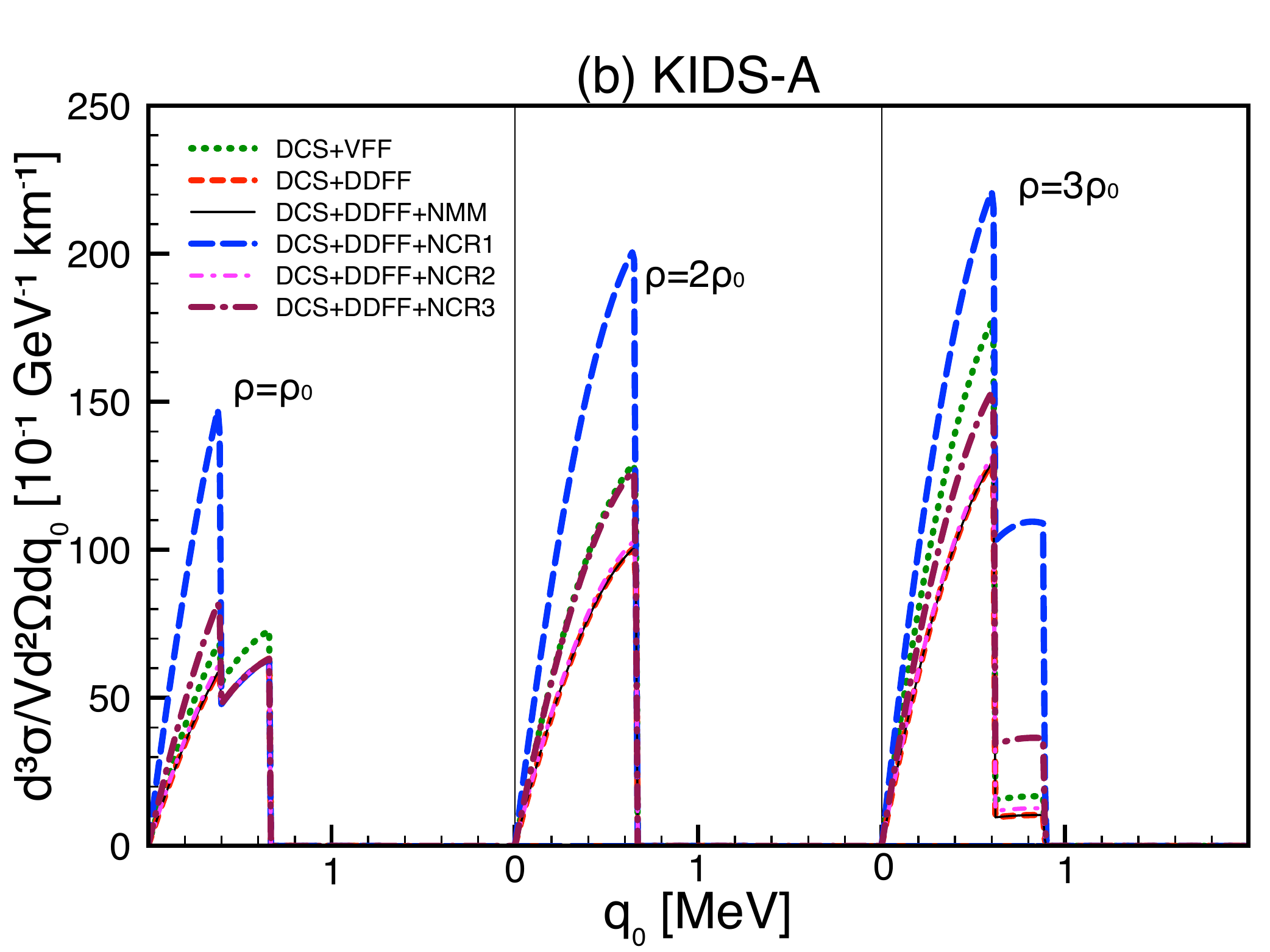}
\\
\includegraphics[width=8cm]{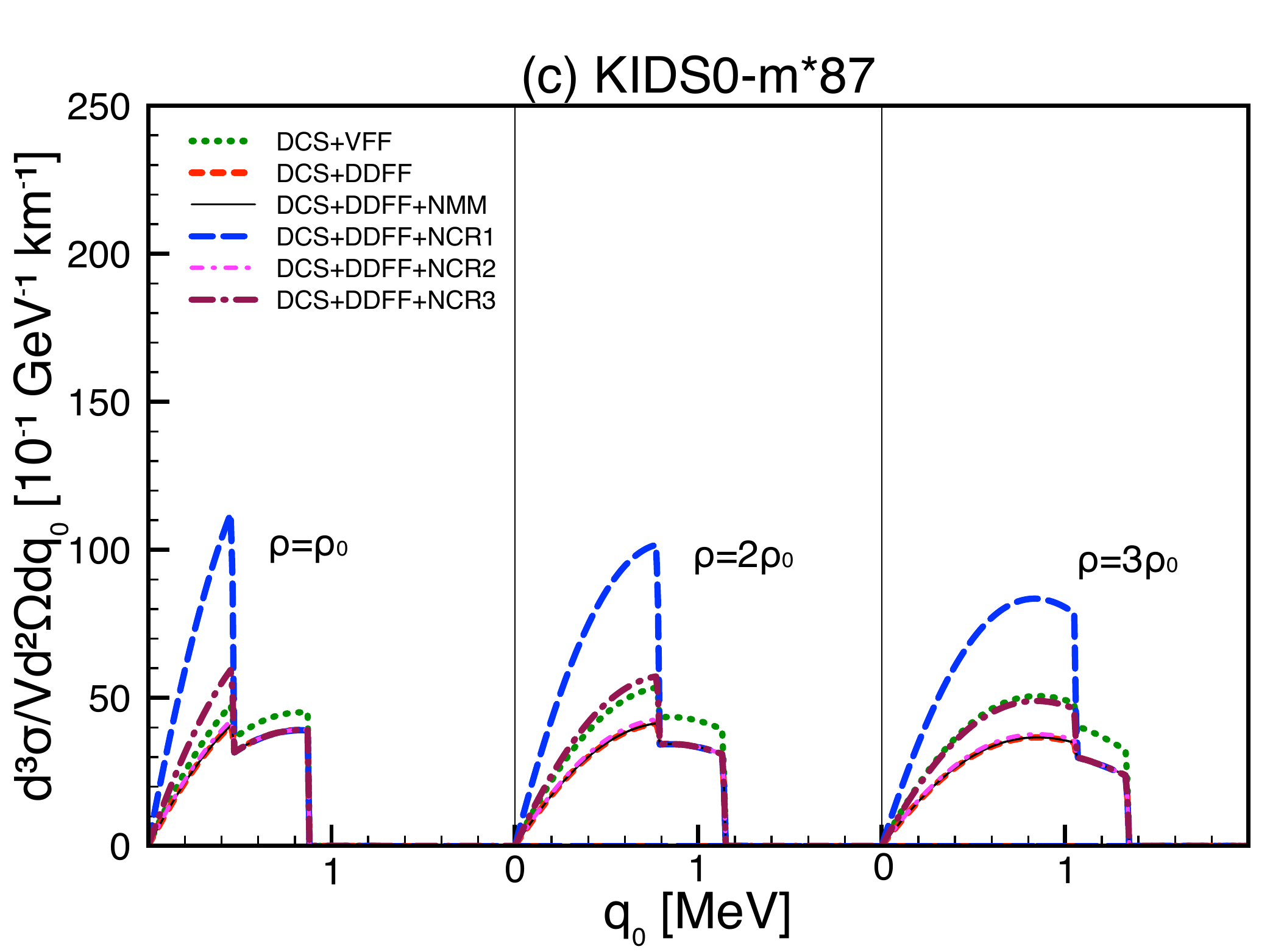}
\includegraphics[width=8cm]{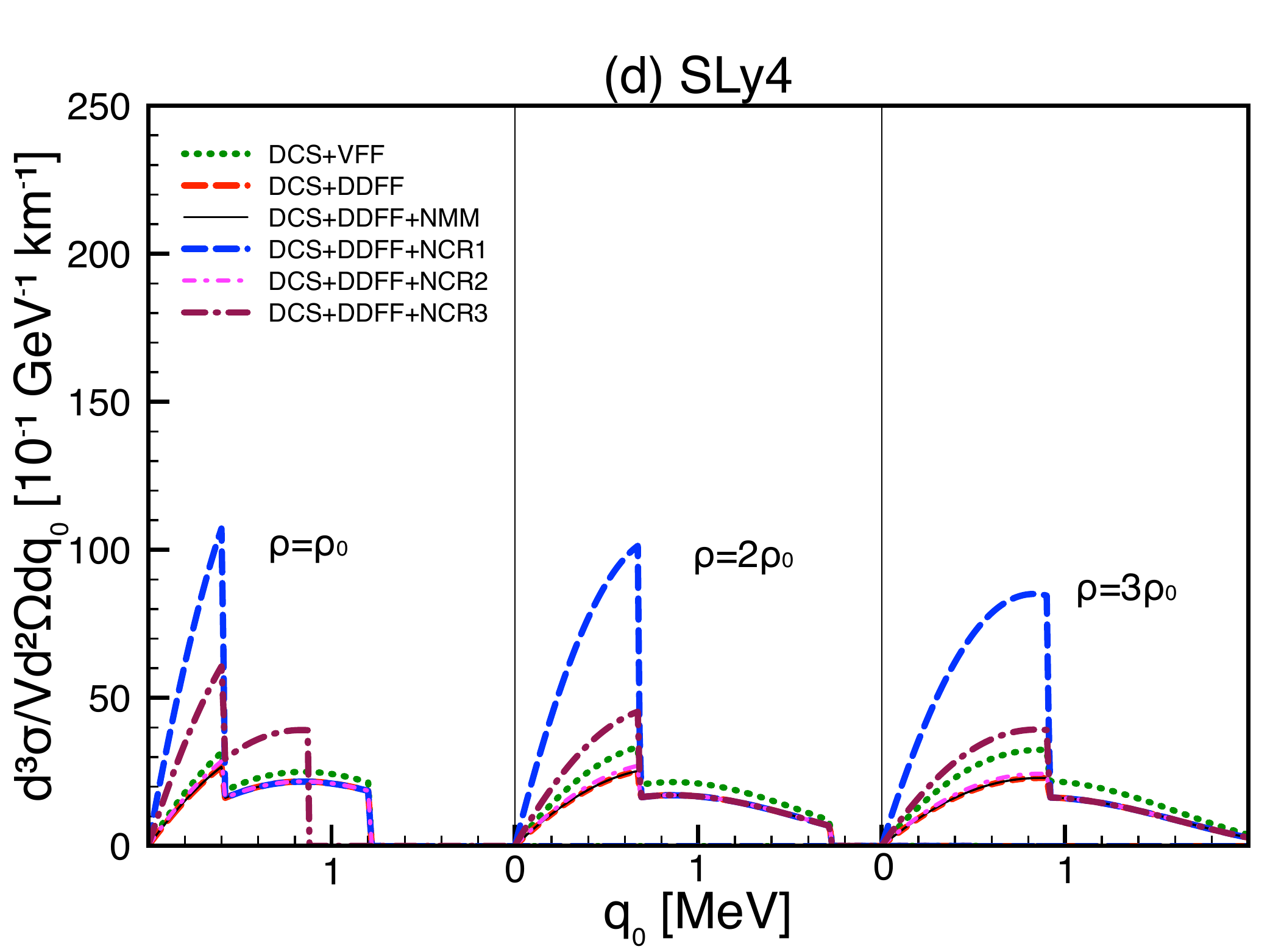}
\end{tabular}
\caption{Triple isotropic differential cross-section (DCS) of the neutrino-nucleon scattering as a function of the transferred energy $q_0$ for $\rho= (1.0-3.0)\,\rho_0$ for the (a) KIDS0, (b) KIDS-A, (c) KIDS0-m*87, and (d) SLy4 models. The numerical results are given with the DCS+VFF (dotted), DCS+DDFF (dashed), DCS+DDFF+NMM (solid), DCS+DDFF+NCR1 (long-dashed), DCS+DDFF+NCR2 (dot-dashed), and DCS+DDFF+NCR3 (dot-long-dashed). Neutrino charge radii are $R_\nu = 3.5$, 1.28 and 2.48 in units of $10^{-5}$ MeV$^{-1}$ for NCR1, NCR2 and NCR3, respectively.}
\label{fig2}
\end{figure}

The temperature of the proto-neutron star that is formed after the bounce of material in the supernova explosion is the order of 10 MeV. It falls down quickly by the deleptonization process and reaches a few MeV in a few minutes. The energy of the neutrino in the $\beta$-equilibrium is at the order of thermal fluctuation, so we choose $E_\nu = 5$ MeV for the neutrinos in the NS throughout the calculation.

The numerical results for the DCS are depicted as a function of $q_0$ for the densities $\rho=\rho_0$ (left), $2\rho_0$ (middle), and $3\rho_0$ (right) in Fig.~\ref{fig2}. In each row, we compute the DCS using different EDF models, i.e., KIDS0, KIDS-A, KIDS0-m*87, and SLy4 from top to bottom. In order to explore the effects of the new ingredients DDFF, NMM, and NCR, we separately show the results with the vacuum form factor (VFF), DDFF, DDFF+NMM, and  DDFF+NCR. Note that the VFF indicates the DDFF at $\rho=0$.  

First, we explain the overall tendency depending on the different EDF models. It is obvious that the DCS from KIDS0 and KIDS-A are qualitatively larger than those from KIDS0-m*87 and SLy4, although quantitative differences are still shown. As discussed in detail in Ref.~\cite{Hutauruk:2022bii}, the larger DCS is originated from the effective neutron mass following the condition $M^*_n\gtrsim M_n$, where $M_n$ is the neutron mass in a vacuum, as in KIDS0 and KIDS-A. This observation can be basically explained by that the DCS is proportional to the $\mathcal{O}(M^{*}_N)$ and $\mathcal{O}(M^{*2}_N)$ terms from the spin summation over the nucleonic tensor given in Eq.~(\ref{eq4}). Physically, the increasing nucleon mass in the scattering process results in decreasing energy transfer $\Delta q_0$ in the $t$ channel at a certain $\sqrt{s}$ value. In turn, the interaction time increases, resulting in a larger cross-section from $\Delta q_0\Delta \tau\gtrsim\hbar/s$, where the $\tau$ stands for the interaction time. The difference between the models becomes more obvious as the density increases as also reported previously in Ref.~\cite{Hutauruk:2022bii}. The different endpoints of each cross-section are determined by the effective nucleon masses depending on the models. 

As for the effects of the DDFF, it turns out that the inclusion of the DDFF provides a non-negligible reduction of the cross-section in comparison to that of VFF. Moreover, the reduction becomes more significant as the density increases. If we compare the maximum values with the DCS+DDFF with those of the DCS+VFF, the reduction rates are 0.83, 0.77, and 0.71 at $\rho=\rho_0$, $2\rho_0$, and $3\rho_0$, respectively, and it is weakly dependent on the EDF models as shown in the figure. Hence, the decreasing behavior of the DCS can be understood in terms of the in-medium behavior of the DDFF, especially $G_A$, as already discussed in the previous subsection.

Now we are in a position to discuss the effects of NMM in the DCS. Here, we used $\mu_\nu = 2.9 \times 10^{-11} \mu_{\rm B}$ for the numerical calculations. This value is constrained from the experiment~\cite{Beda:2013mta} and close to the astronomical observation~\cite{Raffelt:1999gv}. By comparing the DCS+DDFF (dashed) and DCS+DDFF+NMM (solid) in Fig.~\ref{fig2}, 
it is obvious that the NMM gives negligible contributions to the scattering process. Note that the neutrino magnetic tensor currents including the NMM are proportional to $q_0$ as shown in Eq.~(\ref{GAMMAEM}). Hence, the effects of the NMM in the DCS are considerably suppressed in the low $q_0$ region as shown in Fig.~\ref{fig2}, in addition to its extremely small value $\propto10^{-11}\,\mu_B$, in comparison to other scales such as $\mu_n$ and $\mu_p$ in the scattering process.

The NCR relates to the neutrino electric vector current as in Eq.~(\ref{GAMMAEM}) and indicates the EM structure of the particle, although the neutrino has been believed to be a point-like particle in general. In the LAMPF experiment for the measurement $\nu_ee^-\to\nu_ee^-$~\cite{Allen:1992qe}, the NCR was estimated by $R_\nu = 3.5 \times 10^{-5}$ MeV$^{-1}$. In Ref.~\cite{Cadeddu:2018dux} from the new analyses based on the COHERENT elastic neutrino-nucleus scattering, the NCR is given by the upper limit $R_\nu = 1.28 \times 10^{-5}$ MeV$^{-1}$. To explore the tendency of the NCR effect, we also try a middle value $R_\nu = 2.48 \times 10^{-5}$ MeV$^{-1}$. For convenience, we assigned these three NCR values the acronyms NCR1, NCR2, and NCR3, respectively.

Using these values in the numerical calculations, the results are given in the long-dashed (NCR1), dot-dashed (NCR2), and dot-long-dashed (NCR3) lines in the figure. Very interestingly, the DCSs increase sizably with the NCR1~\cite{Allen:1992qe} and NCR3 (between those of Ref.~\cite{Allen:1992qe} and~\cite{Cadeddu:2018dux}) commonly for the different EDF models, while the NCR2~\cite{Cadeddu:2018dux} only makes negligible changes. This observation can be easily understood by that the effects of the NCR2 are about 4 -- 8  times smaller than  NCR1 and NCR3 in the cross-section qualitatively. The effects of the NCR are especially profound for the SLy4 model. We also note that the increase of the DCS due to the NCR is less sensitive to the density than the DDFF since the neutrino EM current is not dependent on the density. The drastic changes observed in the DCS due to the NCR can be understood physically by the spatial extension of the neutrino wave function, increasing the overlap with that for the nucleon, resulting in a rapidly growing DCS.

\subsection{NMFP}
\begin{figure}[tbp]
\begin{tabular}{cc}
\includegraphics[width=8cm]{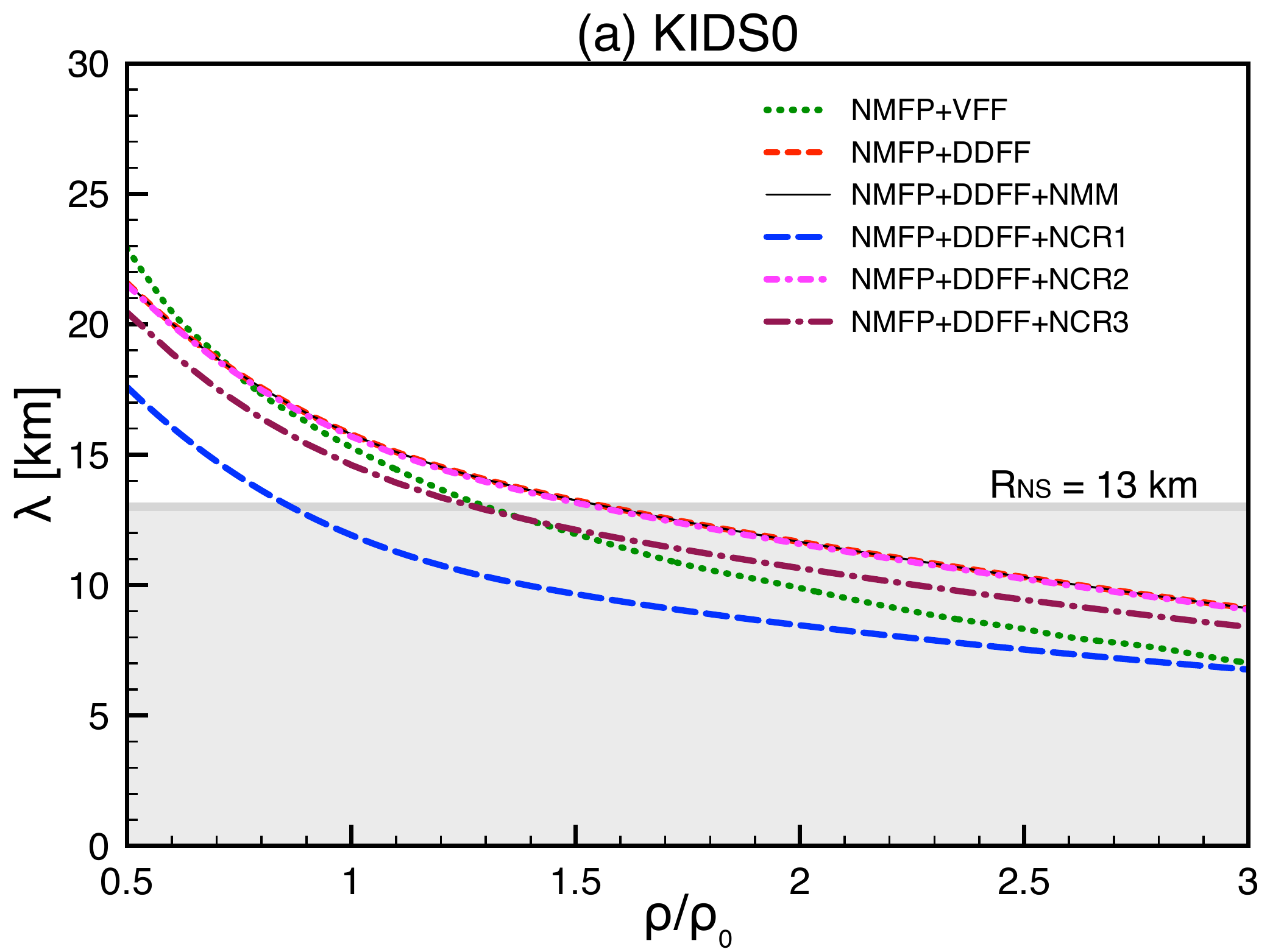}
\includegraphics[width=8cm]{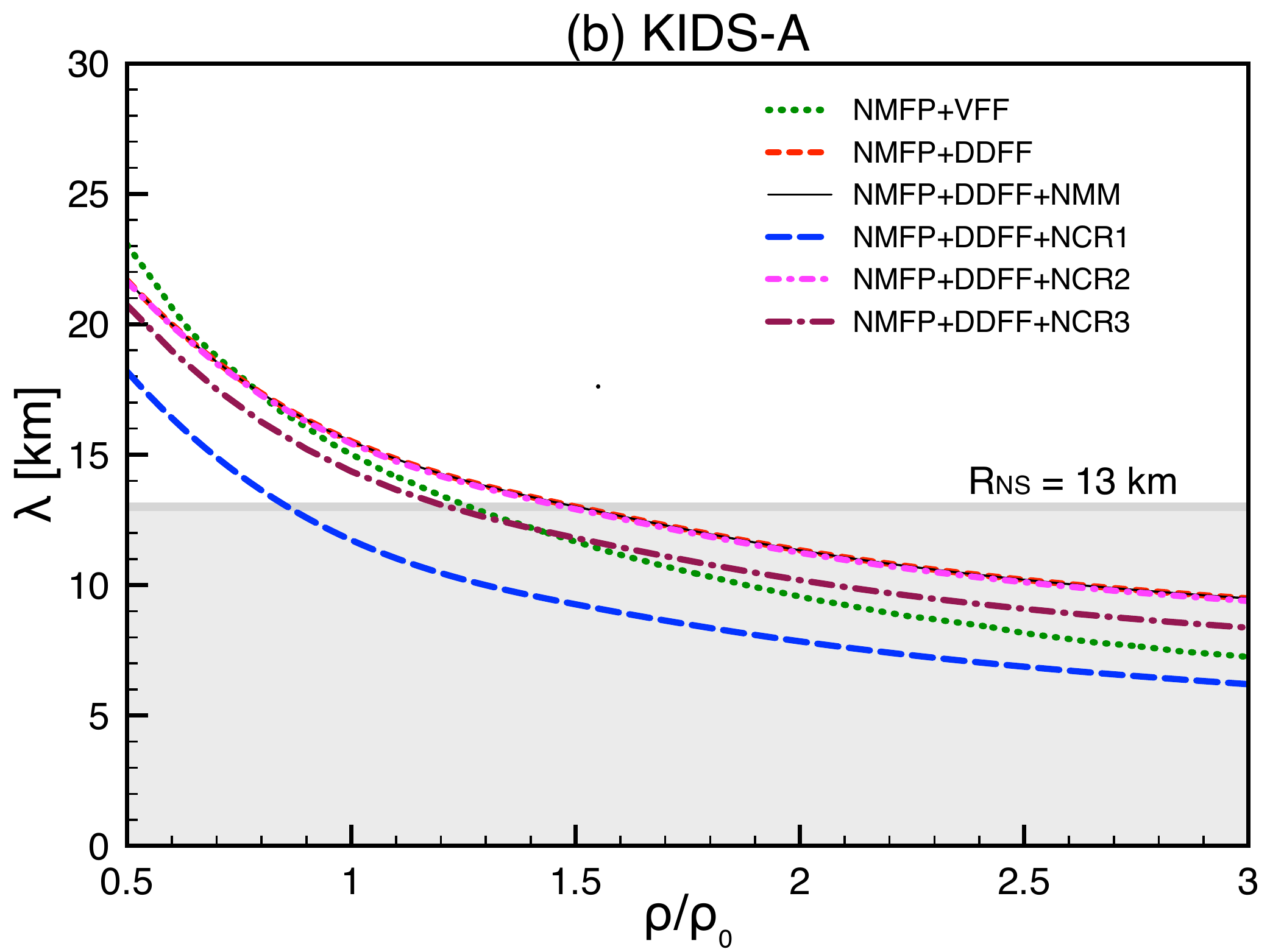}
\\
\includegraphics[width=8cm]{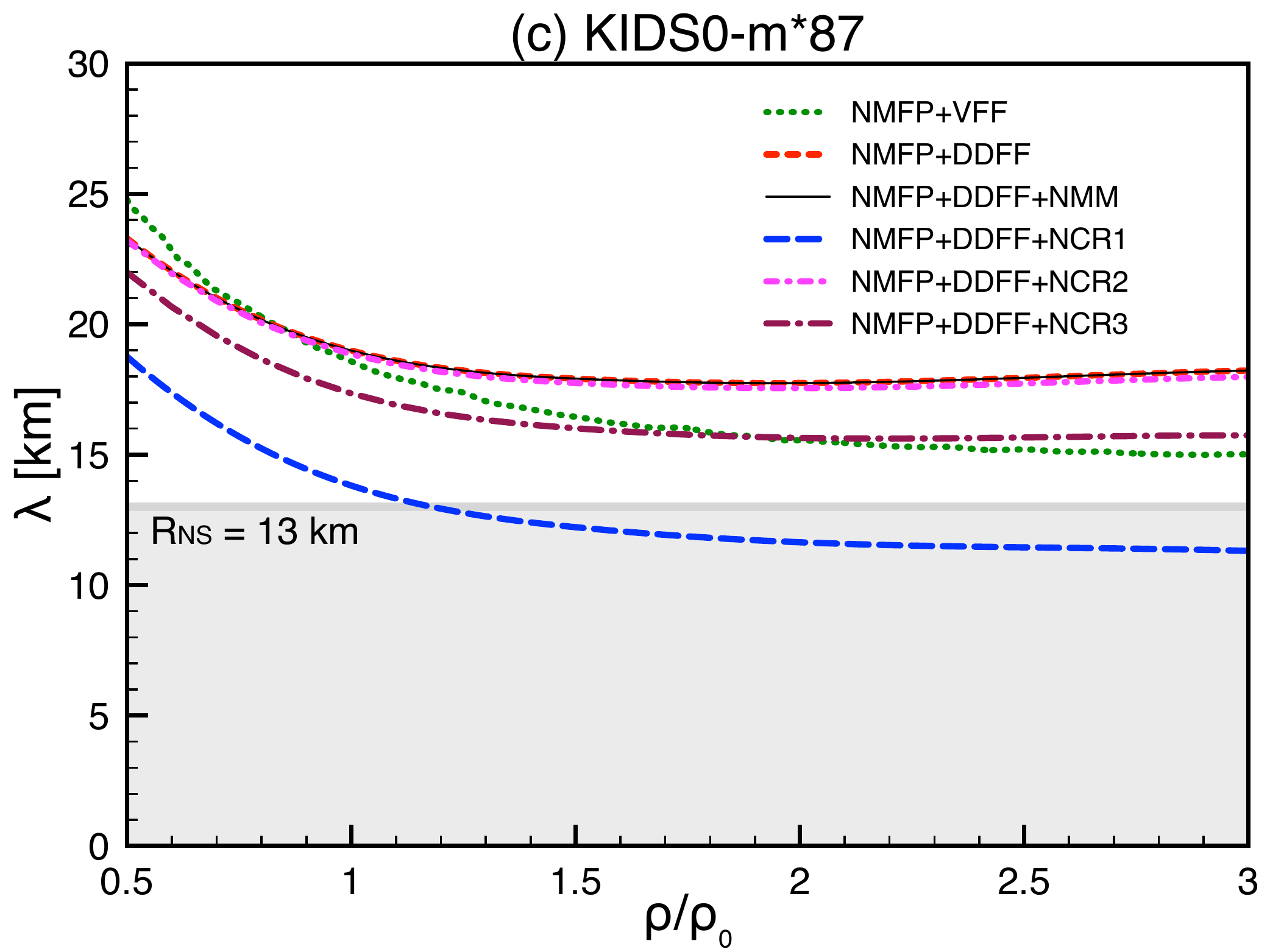}
\includegraphics[width=8cm]{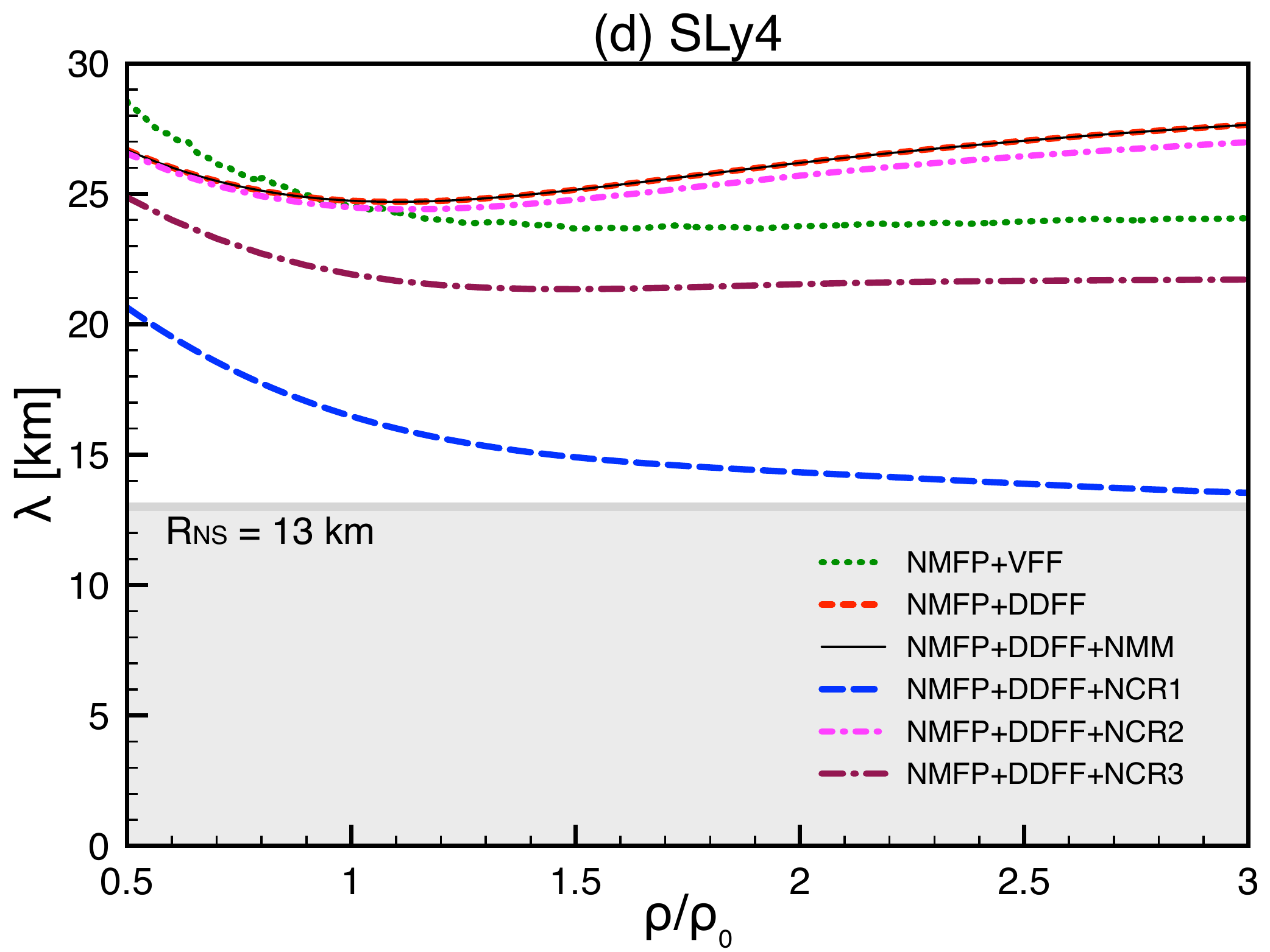}
\\
\includegraphics[width=8cm]{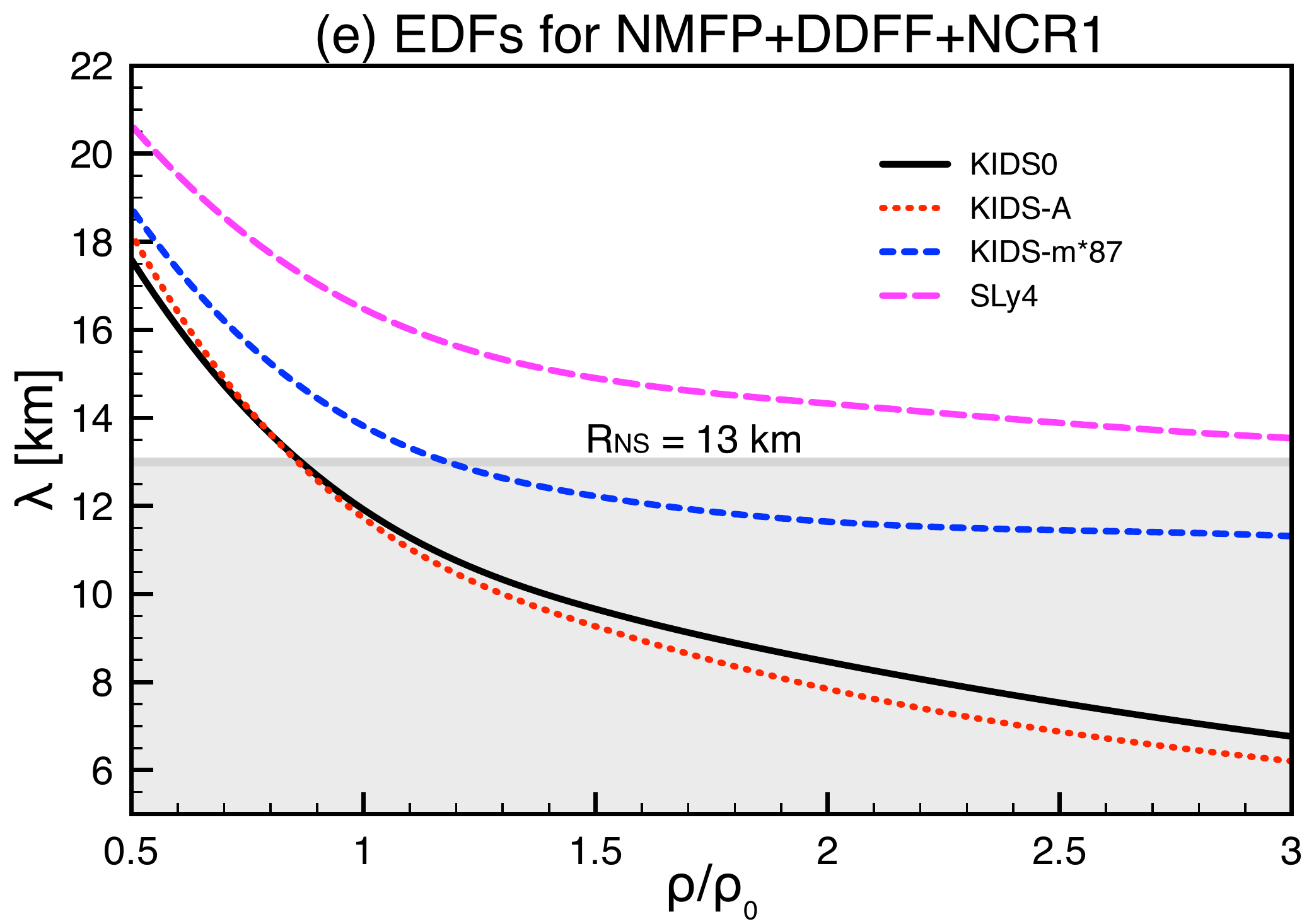}
\end{tabular}
\caption{Neutrino mean free path (NMFP) as a function of density for the (a) KIDS0, (b) KIDS-A, (c) KIDS0-m*87, and (d) SLy4 modes. The numerical results are given with the vacuum form factor NMFP+VFF (dotted), NMFP+DDFF (dashed), NMFP+DDFF+NMM (solid), NMFP+DDFF+NCR1 (long-dashed), NMFP+DDFF+NCR2 (dot-dashed), and NMFP+DDFF+NCR3 (dot-long-dashed). Model dependence is compared with NMFP+DDFF+NCR1 in panel (e). The shaded area stands for the region $R_{\rm NS}=13$~km, an upper bound for the radius of $1.4M_\odot$ mass NS.}
\label{fig3}
\end{figure}

Finally, we present the numerical results for the NMFP in Fig.~\ref{fig3} in the same manner as that for the DCS given above. The NMFP is one of the critical quantities that affect the cooling rate of the NS. It is proportional by construction to the inverse of the DCS, so one can easily expect that shorter NMFP at a certain density for $M^*_n\gtrsim M_n$ in the KIDS0 and KIDS-A models 
in comparison to the others, as shown in Fig.~\ref{fig3}. In common for all the cases, the NMFP is a decreasing function of the density, except for the NMFP+DDFF, NMFP+DDFF+NMM, and NMFP+DDFF+NCR2 in the SLy4 model, being consistent with the consideration that the neutrino-nucleon interaction rate increases in a dense matter. 

In the low-density region, where VFF$\sim$DDFF, the NMFP+VFF, and NMFP+DDFF are close to each other as expected, and the difference gets larger as the density grows for all the cases. Similar to the DCS, the effects of the NCR turn out to be critical throughout the densities with the NCR1 and NCR3 values, decreasing the NMFP by $3-5$ km on average. Again, the NNM does not make any significant contributions to the NMFP, being consistent with the DCS results. Taking into account that the radius of the NS whose mass is $1.4M_\odot$ is $R_{1.4M_\odot} \lesssim 13$~km~\cite{miller2021}, indicated by the horizontal shaded line, as for the results from the KIDS0 and KIDS-A models, the neutrino can not escape freely from the NS if it is emitted at the center of NS where the density is beyond  $\rho\approx1.5 \rho_0$. In contrast, the neutrino seldom experiences weak and EM interactions until it escapes from the NS for the KIDS0-m*87 and SLy4 models.  For instance, there is almost no delay of the neutrino emission for $R_{\rm NS}\approx10$ km because NMFP is always larger than $R_{\rm NS}$.

Interestingly, the NMFP+DDFF for the SLy4 model shows an increasing curve beyond $\rho\approx\rho_0$, being different from others. That obvious difference shown in the SLy4 model is caused by the smallest neutron-effective mass among the EDF models~\cite{Hutauruk:2022bii} in addition to the considerable DDFF effects, which make the DCS reduced. From this observation, the importance of the effective mass is crucial in analyzing the cooling processes of the NS via the neutrinos.  Since the range of the NMFP predicted by the different EDF models is very wide, the application of the present results to the calculation of the thermal evolution of the NS should be necessarily followed up.

Here is a discussion on the effect of symmetry energy. A comparison of the KIDS0 and KIDS-A models clearly shows the role of the symmetry energy because the two models differ evidently in the stiffness of symmetry energy but have similar effective masses. In the results of both DCS and NMFP, the two models show similar behavior up to $3\rho_0$, so the symmetry energy stiffness appears insignificant. Symmetry energy directly affects the particle fraction, giving a larger proton fraction with stiffer symmetry energy. A large proton fraction can ignite the direct URCA process, leading to super-fast cooling of the NS. In addition, it's been shown in \cite{Hutauruk:2022bii} that the particle fraction shows sizable dependence on the symmetry energy at densities higher than $3\rho_0$. Therefore, the effect of the symmetry energy could be probed correctly when the NS cooling is considered explicitly.

\section{Summary and conclusion} \label{sec:summary}
In the present work, we have investigated the neutral-current neutrino-nucleon scattering in the nuclear medium using the quark-meson coupling (QMC) model together with the four different energy-density functional (EDF) models, i.e., KIDS0, KIDS-A, KIDS0-m*87, and SLy4. The nucleon density-dependent form factor (DDFF), differential cross-section (DCS), and neutrino mean free path (NMFP) were computed numerically at various densities. In addition, we also explored the effects of the finite neutrino magnetic moment (NMM) and its EM size via the charge radius (NCR). Below, we list relevant observations found in the present work:
\begin{itemize}
\item Among the weak DDFFs, $G_A$ is a decreasing function of the density, and vice versa for $F^W_2$, whereas $F^W_1$ is almost insensitive to the density in the small $Q^2$ region. These opposite behaviors between $G_A$ and $F^W_2$ can be understood by the different combinations of the density-dependent lower and upper components of the quark spinor in the QMC model. We also find out that the lower part is more sensitive to density and increases with respect to it. The density dependencies in the EM DDFFs turn out to be weak.   
\item The DCS increases with respect to the density in general and is larger in the KIDS0 and KIDS-A models which have $M^*_n\gtrsim M_n$ in comparison to other EDF models. The dominant contribution among the weak DDFFs turns out to be $G_A$, which reduces the DCS with respect to the density as understood by the above discussions. The effect of the NMM is almost negligible since it is highly suppressed in the small $q_0$ region. The finite NCR indicates a larger overlap with the nucleon wave function, resulting in the increase of the DCS in general. The range of increase is, however, strongly dependent on the 
the magnitude of NCR.
\item The NMFP which is inversely proportional to the DCS is scrutinized in the same manner as the DCS. The weak DDFF $G_A$ makes the NMFP increase as understood by its density dependence. 
The inclusion of the NCR can drastically decrease the NMFP by about $5$ km when $R_\nu = 3.5 \times 10^{-5}$ MeV$^{-1}$. 
If we take $R_{\rm NS}\approx 13$ km for instance in the present theoretical framework, the neutrino escapes from the NS almost without interactions up to $\rho\approx3\rho_0$ for $M^*_n\lesssim M_n$. On the contrary for $M^*_n\gtrsim M_n$, DCS increases as density increases, and it leads to NMFP shorter than the NS radius. A decrease in NMFP implies that the interaction of the neutrino with NS matter becomes more probable, and it can impose a non-negligible effect on the thermal evolution of the NS.
\end{itemize}

As discussed previously, there are considerable theoretical uncertainties in the present theoretical framework, depending on the EDF models, density-dependent form factors, and neutrino properties. Differential cross sections and neutrino mean free paths are highly sensitive to the magnitude of NCR, and the effect of NCR is closely coupled to the nuclear physics input, i.e. the effective mass of the nucleon. Hence, it is necessary to investigate the NS thermal evolution to construct a more realistic theoretical model and determine model parameters consistently. Coherent elastic neutrino-nucleus scattering (CE$\nu$NS) is expected to provide brand-new constraints to the beyond-standard-model physics, including NCR and NMM from the laboratory experiment. Recently bounds on NCR and NMM are extracted from the COHERENT data~\cite{jhep2022}. Uncertainty diagnosis in CE$\nu$NS can open a way to obtain more stringent constraints on the relevant quantities and more realistic physical results from the experiment. The effect of effective mass and NCR to CE$\nu$NS cross-section is being investigated. A combination of various independent phenomena such as NS cooling, CE$\nu$NS, quasielastic, and deep inelastic scatterings will help reduce the range of uncertainty. 

Note that, in the present work, we only consider neutrino scattering at zero temperature. It is interesting and challenging to consider the effect of NCR, NMM, and DDFF on the neutrino emissivity of dense matter at finite temperatures. Such a more realistic astrophysical scenario at finite temperatures will be left for our future work.

\section*{Acknowledgments}
This work was supported partially by the National Research Foundation of Korea (NRF) Grants Nos.~2018R1A5A1025563,  2022R1A2C1003964, 2022K2A9A1A0609176, and 2023R1A2C1003177.


\begin{thebibliography}{99}

\bibitem{Hutauruk:2022bii}
P.~T.~P.~Hutauruk, H.~Gil, S.~i.~Nam and C.~H.~Hyun,
\newblock Phys. Rev. C \textbf{106}, no.3, 035802 (2022).

\bibitem{Miller:2007kt}
G.~A.~Miller, E.~Piasetzky and G.~Ron,
\newblock Phys. Rev. Lett. \textbf{101}, 082002 (2008).

\bibitem{Hyde:2004gef}
C.~E.~Hyde and K.~de Jager,
\newblock Ann. Rev. Nucl. Part. Sci. \textbf{54}, 217-267 (2004).

\bibitem{Andivahis:1994rq}
L.~Andivahis \textit{et al.},
\newblock Phys. Rev. D \textbf{50}, 5491-5517 (1994).

\bibitem{Litt:1969my}
J.~Litt  \textit{et al.}, 
\newblock Phys. Lett. B \textbf{31}, 40-44 (1970).

\bibitem{SAMPLE:1997dds}
B.~Mueller \textit{et al.} [SAMPLE],
\newblock Phys. Rev. Lett. \textbf{78}, 3824-3827 (1997).

\bibitem{Horowitz:2003yx}
C.~J.~Horowitz and M.~A.~Perez-Garcia,
\newblock Phys. Rev. C \textbf{68}, 025803 (2003).

\bibitem{Reddy:1998hb}
S.~Reddy, M.~Prakash, J.~M.~Lattimer and J.~A.~Pons,
\newblock Phys. Rev. C \textbf{59}, 2888-2918 (1999).

\bibitem{Sulaksono:2005wv}
A.~Sulaksono, P.~T.~P.~Hutauruk and T.~Mart,
\newblock Phys. Rev. C \textbf{72}, 065801 (2005).

\bibitem{Guo:2020tgx}
G.~Guo, G.~Mart\'\i{}nez-Pinedo, A.~Lohs and T.~Fischer,
\newblock Phys. Rev. D \textbf{102}, no.2, 023037 (2020).

\bibitem{Sulaksono:2006eu}
A.~Sulaksono \textit{et al.}, 
\newblock Phys. Rev. C \textbf{73}, 025803 (2006).

\bibitem{Hutauruk:2006re}
P.~T.~P.~Hutauruk, A.~Sulaksono and T.~Mart,
\newblock Nucl. Phys. A \textbf{782}, 400-405 (2007).

\bibitem{Cloet:2009tx}
I.~C.~Cloet, G.~A.~Miller, E.~Piasetzky and G.~Ron,
\newblock Phys. Rev. Lett. \textbf{103}, 082301 (2009).

\bibitem{Lu:1998tn}
D.~H.~Lu, K.~Tsushima \textit{et al.},
\newblock Phys. Rev. C \textbf{60}, 068201 (1999).

\bibitem{Geesaman:1995yd}
D.~F.~Geesaman, K.~Saito and A.~W.~Thomas,
\newblock Ann. Rev. Nucl. Part. Sci. \textbf{45}, 337-390 (1995).

\bibitem{EuropeanMuon:1983wih}
J.~J.~Aubert \textit{et al.} [European Muon],
\newblock Phys. Lett. B \textbf{123}, 275-278 (1983).

\bibitem{Malace:2008gf}
S.~Malace \textit{et al.} [Jefferson Lab Hall A],
\newblock AIP Conf. Proc. \textbf{1056}, no.1, 141-147 (2008).

\bibitem{Hutauruk:2018cgu}
P.~T.~P.~Hutauruk, Y.~Oh and K.~Tsushima,
\newblock Phys. Rev. D \textbf{98}, no.1, 013009 (2018).

\bibitem{Saito:2005rv}
K.~Saito, K.~Tsushima and A.~W.~Thomas,
\newblock Prog. Part. Nucl. Phys. \textbf{58}, 1-167 (2007).

\bibitem{Hutauruk:2020mhl}
P.~T.~P.~Hutauruk, A.~Sulaksono and K.~Tsushima,
\newblock Nucl. Phys. A \textbf{1017}, 122356 (2022).

\bibitem{plb2009}
C.Y. Ryu, C.H. Hyun, T.-S. Park, and S.W. Hong,
Phys. Lett. B {\bf 674}, 122 (2009).

\bibitem{jpg2010}
C.Y. Ryu, C.H. Hyun, and M.-K. Cheoun,
J. Phys. G {\bf 37}, 105002 (2010).

\bibitem{Super-Kamiokande:2004wqk}
D.~W.~Liu \textit{et al.} [Super-Kamiokande],
\newblock Phys. Rev. Lett. \textbf{93}, 021802 (2004).

\bibitem{TEXONO:2002pra}
H.~B.~Li \textit{et al.} [TEXONO],
\newblock Phys. Rev. Lett. \textbf{90}, 131802 (2003).

\bibitem{MUNU:2003peb}
Z.~Daraktchieva \textit{et al.} [MUNU],
\newblock Phys. Lett. B \textbf{564}, 190-198 (2003).

\bibitem{Beda:2013mta}
A.~G.~Beda \textit{et al.},
\newblock Phys. Part. Nucl. Lett. \textbf{10}, 139-143 (2013).

\bibitem{XENON:2020rca}
E.~Aprile \textit{et al.} [XENON],
\newblock Phys. Rev. D \textbf{102}, no.7, 072004 (2020).

\bibitem{Borexino:2017fbd}
M.~Agostini \textit{et al.} [Borexino],
\newblock Phys. Rev. D \textbf{96}, no.9, 091103 (2017).

\bibitem{Allen:1992qe}
R.~C.~Allen \textit{et al.}, 
\newblock Phys. Rev. D \textbf{47}, 11-28 (1993).

\bibitem{Cadeddu:2018dux}
M.~Cadeddu, C.~Giunti, K.~A.~Kouzakov, Y.~F.~Li, Y.~Y.~Zhang and A.~I.~Studenikin,
Phys. Rev. D \textbf{98}, no.11, 113010 (2018)
[erratum: Phys. Rev. D \textbf{101}, no.5, 059902 (2020)].

\bibitem{CONUS:2022qbb}
H.~Bonet \textit{et al.} [CONUS],
\newblock [arXiv:2201.12257 [hep-ex]].

\bibitem{Jana:2022tsa}
S.~Jana, Y.~P.~Porto-Silva and M.~Sen,
\newblock [arXiv:2203.01950 [hep-ph]].

\bibitem{NOMAD:2009qmu}
V.~Lyubushkin \textit{et al.} [NOMAD],
\newblock Eur. Phys. J. C \textbf{63}, 355-381 (2009).

\bibitem{MiniBooNE:2010xqw}
A.~A.~Aguilar-Arevalo \textit{et al.} [MiniBooNE],
\newblock Phys. Rev. D \textbf{82}, 092005 (2010).

\bibitem{MINERvA:2013kdn}
G.~A.~Fiorentini \textit{et al.} [MINERvA],
\newblock Phys. Rev. Lett. \textbf{111}, 022502 (2013).

\bibitem{Hutauruk:2018qku}
P.~T.~P.~Hutauruk, Y.~Oh and K.~Tsushima,
\newblock Phys. Rev. C \textbf{99}, no.1, 015202 (2019).

\bibitem{Lu:2001mf}
D.~H.~Lu, A.~W.~Thomas and K.~Tsushima,
\newblock [arXiv:nucl-th/0112001 [nucl-th]].

\bibitem{Rakhimov:1998hu}
A.~M.~Rakhimov \textit{et al.}, 
\newblock Nucl. Phys. A \textbf{643}, 383-401 (1998).

\bibitem{Raffelt:1999gv}
G.~G.~Raffelt,
\newblock Phys. Rept. \textbf{320}, 319-327 (1999).

\bibitem{miller2021}
M.~C. Miller {\it et al.}, 
Astrophys. J. {\bf 918}, L28 (2021).

\bibitem{jhep2022}
M.~A. Corona \textit{et al.}, 
JHEP {\bf 09}, 164 (2022).

\end{thebibliography}
\end{document}